\documentclass[epj]{svjour}
\usepackage{float}
\usepackage{tikz}
\usepackage{graphicx}
\usepackage{wrapfig}
\usepackage{amsmath}
\usepackage{lineno}
\usepackage{verbatim}
\usepackage[title]{appendix}
\def\sss{\scriptscriptstyle}

\def\jpsi {{J\mskip -3mu/\mskip -2mu\psi\mskip 2mu}}

\def\makeheadbox{\relax}
\begin{document}
\title{Radiative $b$-baryon decays to measure the photon and $b$-baryon polarization}
\author{Luis Miguel Garc\'ia Mart\'in \inst{1}\thanks{\emph{Luis.Miguel.Garcia@ific.uv.es}}, Brij Jashal\inst{1,2}\thanks{\emph{brij.kishor.jashal@cern.ch
}}, 
Fernando Mart\'inez Vidal\inst{1}\thanks{\emph{Fernando.Martinez@ific.uv.es
}}, 
Arantza Oyanguren\inst{1}\thanks{\emph{Arantza.Oyanguren@ific.uv.es
}}, Shibasis Roy\inst{3,4}\thanks{\emph{shibasisr@imsc.res.in}}  Ria Sain\inst{3,4}\thanks{\emph{riasain@imsc.res.in}} and Rahul Sinha\inst{3,4}\thanks{\emph{sinha@imsc.res.in}}}                     
%
%
\institute{IFIC, Universitat de Val$\grave{e}$ncia-CSIC, Apt. Correus 22085, E-46071 Val$\grave{e}$ncia, Spain \and Tata Institute of Fundamental Research, Mumbai, India \and The Institute of Mathematical
	Sciences, Taramani, Chennai 600113, India \and 
	Homi Bhabha National Institute Training School Complex, Anushakti Nagar, Mumbai 400085, 
	India}
\date{\today}
%
\abstract{The radiative decays of $b$-baryons facilitate the direct measurement of photon helicity in $b\to s\gamma$ transitions thus serving as an important test of physics beyond the Standard Model. In this paper we analyze the complete angular distribution of ground state $b$-baryon ($\Lambda_{b}^{0}$ and $\Xi_{b}^{-}$) radiative decays to multibody final states assuming an initially polarized $b$-baryon sample. Our sensitivity study suggests that the photon polarization asymmetry can be extracted to a good accuracy along with a simultaneous measurement of the initial $b$-baryon polarization. With higher yields of $b$-baryons, achievable in subsequent runs of the Large Hadron Collider (LHC), we find that the photon polarization measurement can play a pivotal role in constraining different new physics scenarios.}

\PACS{{}{IMSc/2019/02/02}}   
%
\authorrunning{.}
\titlerunning{Radiative $b$-baryon decays to measure the photon and $b$-baryon polarization :.}
\maketitle

\section{Introduction}
\label{sec:intro}

The flavour-changing neutral current decays are known to be excellent probes of physics beyond the Standard Model (SM) \cite{Misiak:1992bc,Buras:1993xp,Chetyrkin:1996vx,Kruger:1999xa}. Of particular interest are the loop level Glashow-Iliopoulos-Maiani (GIM) mechanism suppressed processes involving $b\to s \gamma$ transitions that enable the photon polarization to be measured. The helicity flip required for the dipole transition in such decays is sensitive to new physics (NP).
Unfortunately, the  helicity structure of the quark level $b\to s\,\gamma$ transition is hard to analyze using spinless $B$ mesons, as 
handedness of the quark is
difficult to retrieve after hadronization \cite{Grossman:2000rk,Gronau:2001ng,Gronau:2002rz,Hurth:2003vb,Grinstein:2004uu,Becirevic:2012dx,Bishara:2015yta,Kou:2010kn}.

Recently the LHCb experiment performed several indirect measurements concerning
the photon polarization in the $B$ and $B_s$ meson systems. Polarized photons in
$b \to s \,\gamma$ transitions were observed for the first time by analyzing the
up-down asymmetry in $B^+ \to K^+\,\pi^-\,\pi^+\gamma$ decays \cite{Aaij:2014wgo}.
Angular observables in the  $B^{0} \to K^{*0}\,e^+\,e^-$ channel for dielectron
invariant masses of less than 1~GeV$/c^2$, sensitive to the polarization of the
virtual photon, were measured in \cite{Aaij:2015dea}. The first experimental
study of the photon polarization in radiative $B^{0}_{s}$ decays was reported in
\cite{Aaij:2016of} by analyzing the time dependence of the $B^{0}_{s} \to
\phi\gamma$ decay rate.

A large number of $b$-baryons are also being produced at LHCb allowing for
several interesting measurements to be performed using the ground state spin-$1/2$
$b$-baryons. 
The study of spin correlations in radiative decays of $b$-baryons allows one to directly infer the chirality of the dipole transition involved.
The observable of interest, namely, the photon polarization asymmetry $\alpha_{\gamma}$, defined as the ratio of excess left-handed photons over right-handed photons, is well recognized in the context of $b$-baryon radiative decays and  has resulted in several studies \cite{Gremm:1995nx,Mannel:1997xy,Huang:1998ek,Hiller:2001zj,Aliev:2005np,He:2006ud,Leitner:2006nb,Legger:2006cq,Oliver:2010im,Liu:2011ema,Blake:2015tda}. Despite the rich phenomenology, the photon polarization asymmetry in radiative $b$-baryon decays is yet to be explored in experiments.
In addition, it is worth mentioning that the initial $b$-baryon can have non-zero transverse polarization \cite{Leader:1996hm,Dharmaratna:1996xd} depending on the various production mechanism of a $b$-quark at the Large Hadron Collider (LHC). For finite quark masses ($m_{b}\neq 0$), a $b$-quark produced in a QCD process can have transverse polarization of the order of 10\% \cite{Dharmaratna:1996xd,Galanti:2015pqa}. During hadronization, the polarization of the $b$-quark is retained to a large extent by the $b$-baryon as the depolarization effects \cite{Dharmaratna:1996xd} due to QCD interactions are suppressed by a factor of $\Lambda_{\text{QCD}}$/$m_{b}$, a feature common to all $b$-baryons. A systematic study of $b$-baryon polarization $P_{b}$ \cite{Aaij:2013oxa,CMS:2016iaf}, defined as the production asymmetry between its up and down spin components, is therefore important for our understanding of the production mechanism and hadronization process of heavy quarks. 

In this paper we revisit the $\Lambda_{b}^{0}\to \Lambda^{0} \,\gamma$ and the 
analogous $\Xi_b^{-} \to \Xi^{-} \gamma$ decays. 
The $\Lambda^0$ and $\Xi^-$ subsequently decay to multibody final states that provide powerful handles not only on the photon polarization but also the initial $b$-baryon polarization. 
The complete angular distribution of the decay chains
$\Lambda_{b}^{0}\to \Lambda^0 \,\gamma$ and $\Xi_{b}^-\to \Xi^-(\to \Lambda^0\,\pi^-)\,\gamma$, with the $\Lambda^0$ decaying into $p\, \pi^-$, is a function of the photon polarization, $\alpha_{\gamma}$, the initial $b$-baryon polarization, $P_b$, and known decay asymmetry parameters of intermediate baryons. We explore the  potential of simultaneous measurement of the photon and $b$-baryon polarizations at LHCb through the study of the angular distribution of $b$-baryon decays using Monte Carlo simulations. With the expected yield from the LHC Run II a sensitivity of about 0.15 is achievable for the photon polarization along with a measurement of  $\Lambda_{b}^{0}$ polarization with a precision better than 10\%. Despite of the challenges in $\Xi_{b}^{-}$ due to scarcity of data, the decay mode $\Xi_{b}^{-}\to \Xi^{-}\gamma$ remains a promising channel for Run II and beyond where a sensitivity of the order of 0.2 can be achieved in $\alpha_{\gamma}$ measurement along with a first time measurement of $\Xi_{b}$ polarization.  

The paper is laid out as follows: in Sec.~\ref{sec:ang} we have discussed the angular distribution of the radiative decays of $\Lambda_{b}^0$ and $\Xi_{b}^-$. Sec.~\ref{sec:exp} is devoted to the sensitivity analysis of these rare decays at the LHCb experiment. Finally, we provide constraints on new physics scenarios from the photon polarization asymmetry in Sec.~\ref{sec:flavio} before concluding in Sec.~\ref{sec:conc}.
\section{Angular distribution of radiative $b$-baryon decays}
\label{sec:ang}
\subsection{Case study of $\Lambda_{b}^{0} \to \Lambda^0 \,\gamma$}\label{subsec:Lb}
\begin{figure}
\centering
\resizebox{\columnwidth}{!}{\includegraphics[trim={1cm 14cm 2cm 2cm}, clip=true,scale=1, ]{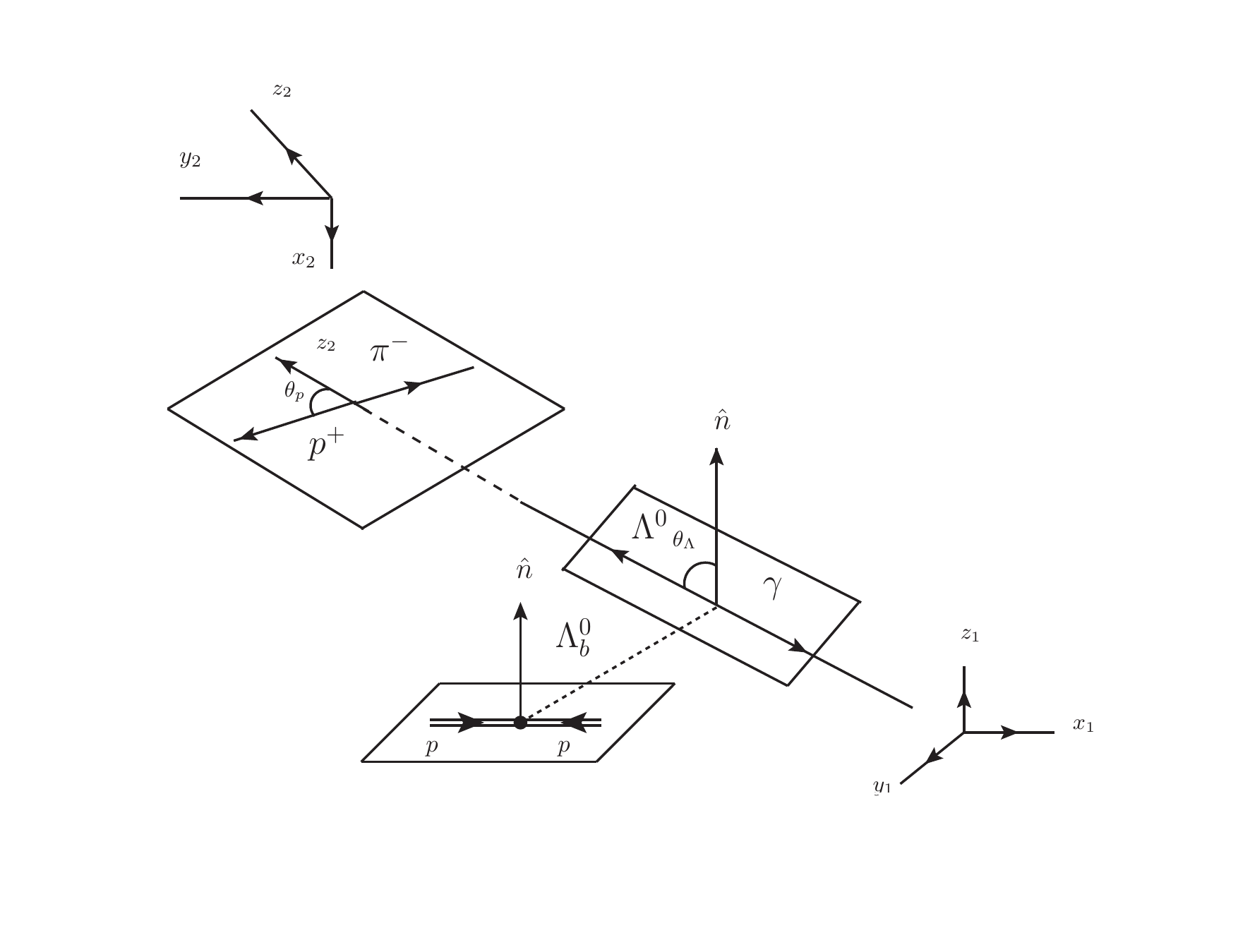}}
\caption{Schematic view of the $\Lambda_{b}^{0} \to \Lambda^0 \,\gamma$ decay.}
\label{fig:Lb_Decay}
\end{figure}

Assuming an initially polarized sample of $\Lambda_{b}$ we study the complete 
angular distribution of $\Lambda_{b}^{0} \to \Lambda^0 \,\gamma$ where the $\Lambda$ 
subsequently decays to a proton ($p$) and pion ($\pi^{-}$).
The transverse $\Lambda_{b}^{0}$ polarization is defined \cite{Aaij:2013oxa,CMS:2016iaf} as the mean value of the $\Lambda_{b}^{0}$ spin along the unit vector \begin{eqnarray}
\hat{n}=\frac{\overrightarrow{p}_{p}\times 
\overrightarrow{p}_{\Lambda_{b}}}{\vert \overrightarrow{p}_{p}\times 
\overrightarrow{p}_{\Lambda_{b}} \vert},
\end{eqnarray}  
normal to the production plane, where $\overrightarrow{p}_{p}$ is the 
counterclockwise proton beam direction and $\overrightarrow{p}_{\Lambda_{b}}$ 
the $\Lambda_{b}^{0}$ momentum.
A schematic view of the decay is shown in Fig.~\ref{fig:Lb_Decay}.
The primary decay of $\Lambda_{b}^0\to \Lambda^0 \,
\gamma$ is described by two complex helicity amplitudes 
$H_{\lambda_{\Lambda},\lambda_{\gamma}}$ with $\lambda_{\Lambda}=\pm1/2$, 
$\lambda_{\gamma}=\pm 1$ referring to the helicity values of the respective 
particles \cite{Gutsche:2013pp}. To define the angles we consider two Cartesian coordinate systems 
attached to the rest frames of $\Lambda_{b}^{0}$ and $\Lambda^0$. In the rest frame 
of $\Lambda_{b}^{0}$, the $z_{1}$-axis is chosen to be parallel to the $\Lambda^0_{b}$ 
polarization direction $\hat{n}$. The relevant kinematic variables are defined as follows:
\begin{itemize}
\item $\theta_{\Lambda}$ is the polar angle of the $\Lambda^{0}$ momentum relative to 
$\hat{n}$ in $\Lambda_{b}^0$ rest frame, $0\leq\theta_{\Lambda}<\pi$. 
\item $\theta_{p},\phi_{p}$ are the polar and azimuthal angle of the proton 
momentum defined with respect to the axes $z_{2}=\overrightarrow{p}_{\Lambda}/\vert 
\overrightarrow{p}_{\Lambda} \vert$ and 
$y_{2}=\hat{n}\times\overrightarrow{p}_{\Lambda}/\vert 
\hat{n}\times\overrightarrow{p}_{\Lambda} \vert$ in $\Lambda^0$ rest frame.
\end{itemize}

 The angular distribution is:
\begin{flushleft}
\begin{align}\label{eq:Lb_AngDist}
 W(\theta_{\Lambda},\theta_{p}) \propto & \,  1-\alpha_{\Lambda}P_{\Lambda_b} \cos{\theta_p}\cos{\theta_{\Lambda}} \nonumber \\
&-\alpha_{\gamma}(\alpha_{\Lambda} \cos{\theta_p}-P_{\Lambda_b}\cos{\theta_{\Lambda}})
\end{align}
\end{flushleft} 

Here, $P_{\Lambda_b}$ is the initial $\Lambda_b$ polarization and $\alpha_\Lambda$ is the $\Lambda^0$ weak decay parameter, defined in Table \ref{tab:dec param}.  

\subsection{Case study of $\Xi_{b}^{-} \to \Xi^{-} \gamma$}\label{subsec:Xib}
\begin{figure}
\centering
\resizebox{\columnwidth}{!}{\includegraphics[trim={2.8cm 18.5cm 3cm 3cm}, clip=true,scale=1, ]{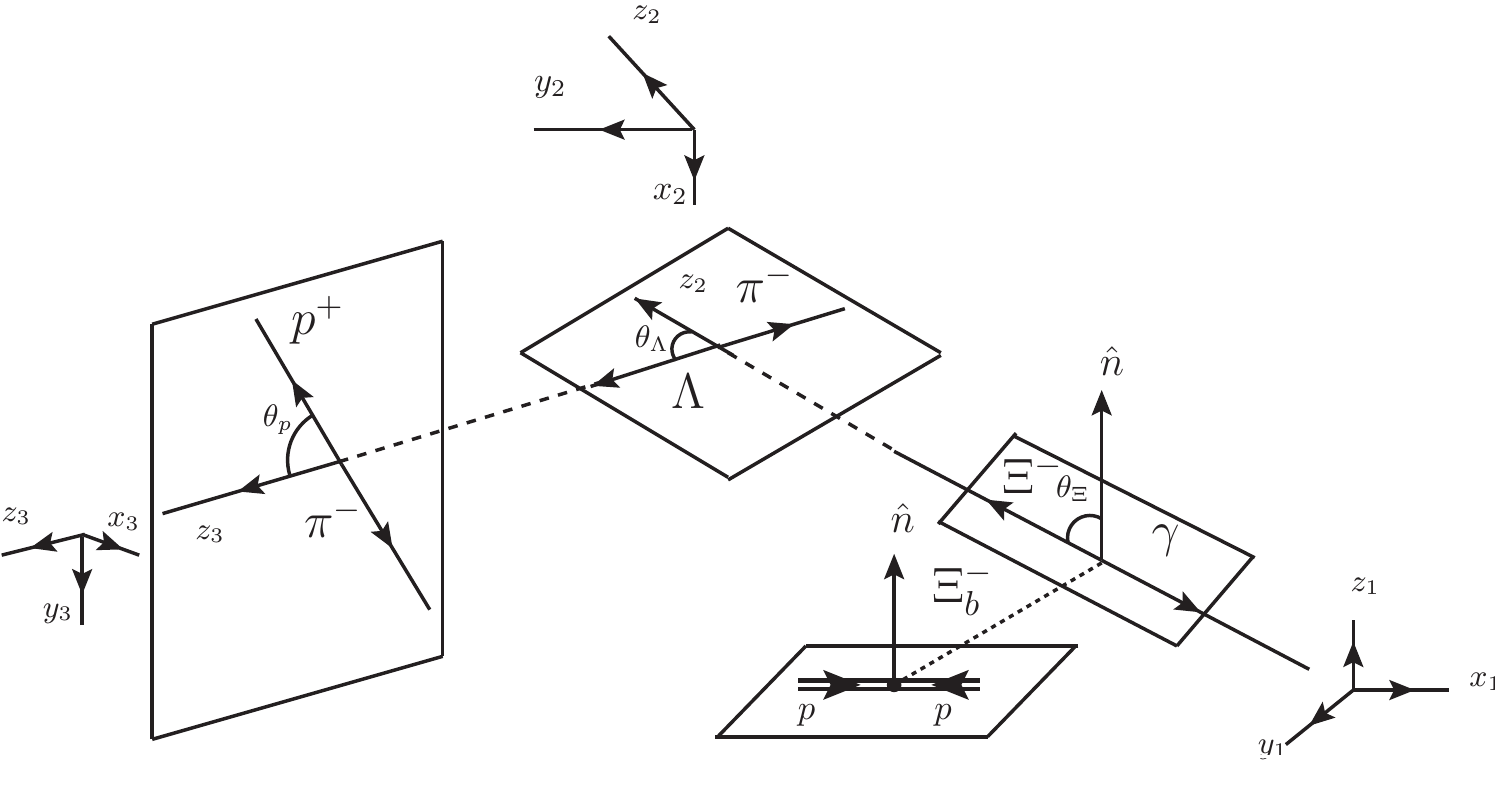}}
\caption{Schematic view of the $\Xi_b^{-} \to \Xi^{-} \gamma$ decay.}
\label{fig:Xib_Decay}
\end{figure}

Assuming an initially polarized sample of $\Xi_{b}^{-}$ we study the complete 
angular distribution of $\Xi_{b}^{-} \to \Xi^{-} \gamma$, considering the subsequent 
decay $\Xi^{-} \to \Lambda^0 \, \pi^-$. The $\Lambda^0$ further decays to a proton and pion, resulting in a $p \,\pi^{-}\,\pi^{-}\,\gamma$ final state.
The transverse polarization of $\Xi_{b}^-$ is defined as the mean value of 
$\Xi_{b}^-$ spin along the unit vector \begin{eqnarray}
\hat{n}=\frac{\overrightarrow{p}_{p}\times \overrightarrow{p}_{\Xi_{b}}}{\vert 
\overrightarrow{p}_{p}\times \overrightarrow{p}_{\Xi_{b}} \vert},
\end{eqnarray}  
normal to the production plane, where $\overrightarrow{p}_{p}$ is the 
counter-clockwise proton beam direction and $ \overrightarrow{p}_{\Xi_{b}}$ the 
$\Xi_{b}^-$ momentum. 
A schematic view of the decay is shown in Fig. \ref{fig:Xib_Decay}.
The primary decay $\Xi_{b}^-\to \Xi^-\, \gamma$ is described 
by two complex helicity amplitudes $H_{\lambda_{\Xi},\lambda_{\jpsi}}$ with 
$\lambda_{\Xi}=\pm1/2$, $\lambda_{\gamma}=\pm 1$ referring to the helicity 
values of the respective particles. To define the angles we consider three 
Cartesian coordinate systems attached to the rest frames of $\Xi_{b}^-$, $\Xi^-$ 
and $\Lambda^0$.  In the rest frame 
of $\Xi_{b}^{-}$, the $z_{1}$-axis is chosen to be parallel to the $\Xi^{-}_{b}$ 
polarization direction $\hat{n}$. Similar to the previous case, the relevant kinematic variables are defined as follows:
 \begin{itemize}
\item $\theta_{\Xi}$ is the polar angle of the $\Xi^{-}$ momentum relative to $\hat{n}$ 
in $\Xi_{b}^{-}$ rest frame, $0\leq\theta_{\Xi}<\pi$. 
\item $\theta_{\Lambda},\phi_{\Lambda}$ are the polar and azimuthal angle of the $\Lambda^0$ 
momentum defined with respect to the axes $z_{2}=\overrightarrow{p}_{\Xi}/\vert 
\overrightarrow{p}_{\Xi} \vert$ and 
$y_{2}=\hat{n}\times\overrightarrow{p}_{\Xi}/\vert 
\hat{n}\times\overrightarrow{p}_{\Xi} \vert$ in $\Xi$ rest frame.
\item $\theta_{p},\phi_{p}$ are the polar and azimuthal angle of the proton 
momentum defined with respect to the axes $z_{3}=\overrightarrow{p}_{\Lambda}/\vert 
\overrightarrow{p}_{\Lambda} \vert$ and 
$y_{3}=\hat{n}\times\overrightarrow{p}_{\Lambda}/\vert 
\hat{n}\times\overrightarrow{p}_{\Lambda} \vert$ in $\Lambda^0$ rest frame.
\end
{itemize}
The complete angular distribution reads,
\begin{flushleft}
\begin{align}\label{eq:Xi_AngDist}
& W(\eta,\theta_{\Lambda},\theta_{p},\theta_{\Xi})\propto 1+ 
\alpha_{\Lambda}\alpha_{\Xi}\cos \theta_{p} +\alpha_{\gamma}\alpha_{\Xi}\cos 
\theta_{\Lambda} \nonumber \\  & +\alpha_{\Lambda}\alpha_{\gamma}\cos \theta_{p} \cos 
\theta_{\Lambda}-2 \alpha_{\Lambda}\alpha_{\gamma} \rm{Re}(e^{i\eta }z_{\Xi})\sin 
\theta_{p} \sin \theta_{\Lambda} \nonumber \\  & - P_{\Xi_{b}}\alpha_{\Xi}\cos \theta_{\Xi} 
\cos \theta_{\Lambda} -P_{\Xi_{b}}\alpha_{\gamma}\cos \theta_{\Xi} \\  
&-P_{\Xi_{b}}\alpha_{\Xi}\alpha_{\Lambda}\alpha_{\gamma}\cos \theta_{\Xi} \cos 
\theta_{p}  -P_{\Xi_{b}}\alpha_{\Lambda}\cos \theta_{\Xi} \cos \theta_{\Lambda} \cos 
\theta_{p} \nonumber\\&+2 \alpha_{\Lambda} P_{\Xi_{b}} \rm{Re}(e^{i\eta 
}z_{\Xi})\cos \theta_{\Xi} \sin \theta_{p} \sin \theta_{\Lambda}, \nonumber
\end{align}
\end{flushleft}
where $\eta=\phi_{\Lambda}-\phi_{p}$, and the $\Xi_{b}^-$ polarization matrix 
$\rho_{\sss\lambda_{\sss\Xi_{b}},\sss\lambda_{\sss\Xi_{b}}^{\prime} }$ is 
defined 
as 
$\rho_{\sss\frac{1}{2},\sss\frac{1}{2}}+\rho_{-\sss\frac{1}{2},-\sss\frac{1}{2}}=1$ and  $\text{Tr}[\rho\hat{n}]= (\rho_{\sss\frac{1}{2},\sss\frac{1}{2}}-\rho_{-\sss\frac{1}{2},-\sss\frac{1}{2}})\, \hat{z}=P_{\Xi_{b}}\, \hat{z} $, 
with the off-diagonal entries averaging out to zero in absence of correlation 
between production and decay mechanisms.
Here $P_{\Xi_b}$ is the initial $\Xi_b^-$ polarization, and $\alpha_{\Xi}$ is the $\Xi^-$ weak decay parameter. 
The complete expressions of decay parameters and their values are
summarized in Table \ref{tab:dec param}.
 \begin{table}[H]
\resizebox{\columnwidth}{!}{
\begin{tabular}{c c c}
\hline \noalign{\smallskip}
    &  Expression  & Obs. Values \\
    \noalign{\smallskip}
 \hline
 \noalign{\smallskip}
 $\alpha_{\gamma}$ & $\displaystyle\frac{\vert 
H_{\lambda_{\Xi}= -1/2,\lambda_{\gamma}=-1} 
\vert^2-\vert H_{\lambda_{\Xi}=1/2,\lambda_{\gamma}=1} \vert^2}{\vert 
H_{\lambda_{\Xi}=-1/2,\lambda_{\gamma}=-1} \vert^2+\vert 
H_{\lambda_{\Xi}=1/2,\lambda_{\gamma}=1} \vert^2}$ & -\\[3ex]
\hline
\noalign{\smallskip}
 $\alpha_{\Xi}$ & $\displaystyle\frac{\vert H_{\lambda_{\Lambda}=1/2} \vert^2-\vert H_{\lambda_{\Lambda}=-1/2} \vert^2}{\vert H_{\lambda_{\Lambda}=1/2} \vert^2+\vert H_{\lambda_{\Lambda}=-1/2} \vert^2}$ & $-0.458\pm 0.012$ \cite{PDG18} \\[3ex]
\hline
\noalign{\smallskip}
 $\alpha_{\Lambda}$ & $\displaystyle\frac{\vert H_{\lambda_{p}=1/2} \vert^2-\vert H_{\lambda_{p}=-1/2} \vert^2}{\vert H_{\lambda_{p}=1/2} \vert^2+\vert H_{\lambda_{p}=-1/2} \vert^2}$ & $0.642\pm0.013$ \cite{PDG18}
 \\[3ex]
\hline
\noalign{\smallskip}
  $z_{\Xi}$ & $\displaystyle\frac{ H_{\lambda_{\Lambda}=1/2}^{*}  H_{\lambda_{\Lambda}=-1/2} }{\vert H_{\lambda_{\Lambda}=1/2} \vert^2+\vert H_{\lambda_{\Lambda}=-1/2} \vert^2}$&-
  \\[3ex]
\hline 
\hline
 \end{tabular} 
 }
 \caption{Definition and measured values of the decay parameters and photon polarization used in Eq. \ref{eq:Lb_AngDist} and Eq. \ref{eq:Xi_AngDist} }\label{tab:dec param}
 \end{table}

\vspace{0.5cm}


\section{Experimental prospects at the LHCb experiment}
\label{sec:exp}

The potential to measure the photon and $b$-baryon polarization at LHCb through the angular distribution of $b$-baryon decays can be studied using Monte Carlo simulations. 
The expected sensitivity as a function of the values of the $\alpha_\gamma$, $P_{\Lambda_b}$ and $P_{\Xi_b}$ parameters, and depending on the number of reconstructed events, is determined. Effects due to the angular acceptance and resolution, coming from the event reconstruction, are included in the simulations. The impact of the amount and shape of the possible background is also studied. Results are detailed along this section. Since the expected number of events depend on the $b$-hadron production fractions in proton-proton collisions at the LHC energy, a summary of the present status is also discussed.  

\subsection{$b$-baryon fragmentation fractions}
\label{sec:fragmentation}
There are various processes through which a $b$ quark can be produced in LHC. The $b$ quark can then combine with a light diquark to produce a $b$-baryon. The probability of a $b$-quark fragment into a baryon is given by the fragmentation fraction ($f_{B}$).
It is customary to express the fragmentation fraction
\cite{Galanti:2015pqa,Aaij:2014jyk,Hsiao:2015txa,Jiang:2018iqa} for a specific 
$b$-baryon as  
\begin{eqnarray}
R_{B}=\frac{f_{B}}{f_{i}+f_{j}},
\end{eqnarray}
where $(f_{i}, \,f_{j})$ are: $(f_{u},\,f_{d})$ for $B=\Lambda_{b}^{0}$,  
$(f_{u},\,f_{s})$ for $B=\Xi_{b}^{0}$, $(f_{d},\,f_{s})$ for 
 $B=\Xi_{b}^{-}$ and $f_{u,d,s}\equiv {\cal B}(b\to B^{-},\, \bar{B}^{0}, \, \bar{B}^{0}_{s})$. $R_{B}$ is experimentally measured \cite{Aaij:2011jp,Amhis:2016xyh} and 
 along with the knowledge of $f_{u,s,d}$ one can estimate $f_{B}$ \cite{Aaij:2014jyk,Aaij:2013qqa,Aaij:2016avz}. It is important to note that $f_{\Lambda_{b}}/(f_{u}+f_{d})$ has a dependence on 
 $p_{\rm{T}}$ of the final state particles 
 \cite{Aaij:2014jyk,Aaij:2011jp,Aaij:2015fea}. 
  Latest results from LHCb yields \cite{Aaij:2011jp}
 \begin{eqnarray}
\frac{ f_{\Lambda_{b}}}{f_{u}+f_{d}}=  0.404\pm 0.017\pm0.027\pm0.105.
\end{eqnarray}
 Recently LHCb has measured the fragmentation fraction of a $b$ quark hadronizing into a $\Xi_{b}^{-}$ ($f_{\Xi_{b}^{-}}$) with respect to fragmentation fraction of a $b$ quark hadronizing into a $\Lambda_{b}$ ($f_{\Lambda_{b}}$), using the $\Xi_{b}^{-}\to \Xi^{-} J/\psi$ and $\Lambda_{b}\to \Lambda J/\psi$ decay modes \cite{Aaij:2019ezy,Voloshin:2015xxa}. 

Using the data samples collected at $\sqrt{s}=7,\, 8, \,\text{and}\, 13$ TeV  to
measure the ratio of production rates of $\Xi_{b}^{-}$ to $\Lambda_{b}^0$ , with pseudorapidity and $p_{T}$ in the range $2\leq\eta\leq 6$, $p_{T}\leq \;20 \text{GeV}/c$, the following values are observed:

\begin{align}
\label{eq:fragm_b-baryons}
& \frac{f_{\Xi_{b}^{-}}}{f_{\Lambda_{b}}}=(6.7 \pm 2.1)\times 10^{-2} \quad \nonumber [\sqrt{s}=\text{7,\;{\rm and} 8 TeV}],\\
& \frac{f_{\Xi_{b}^{-}}}{f_{\Lambda_{b}}}=(8.2\pm 2.6)\times 10^{-2} \quad [\sqrt{s}=\text{13 TeV}]. 
\end{align}
 In these results the uncertainty is dominated by the departure from the SU(3) relation $\Gamma(\Xi_{b}^{-} \to \Xi^{-} \, J/\psi) = \frac{2}{3} $ $\Gamma(\Lambda_{b}^{0} \to \Lambda^0 \, J/\psi)$, coming from $SU(3)$ breaking of the order of 30\% \cite{Aaij:2019ezy}.

\begin{figure}
\centering
\resizebox{0.4\textwidth}{!}{\includegraphics{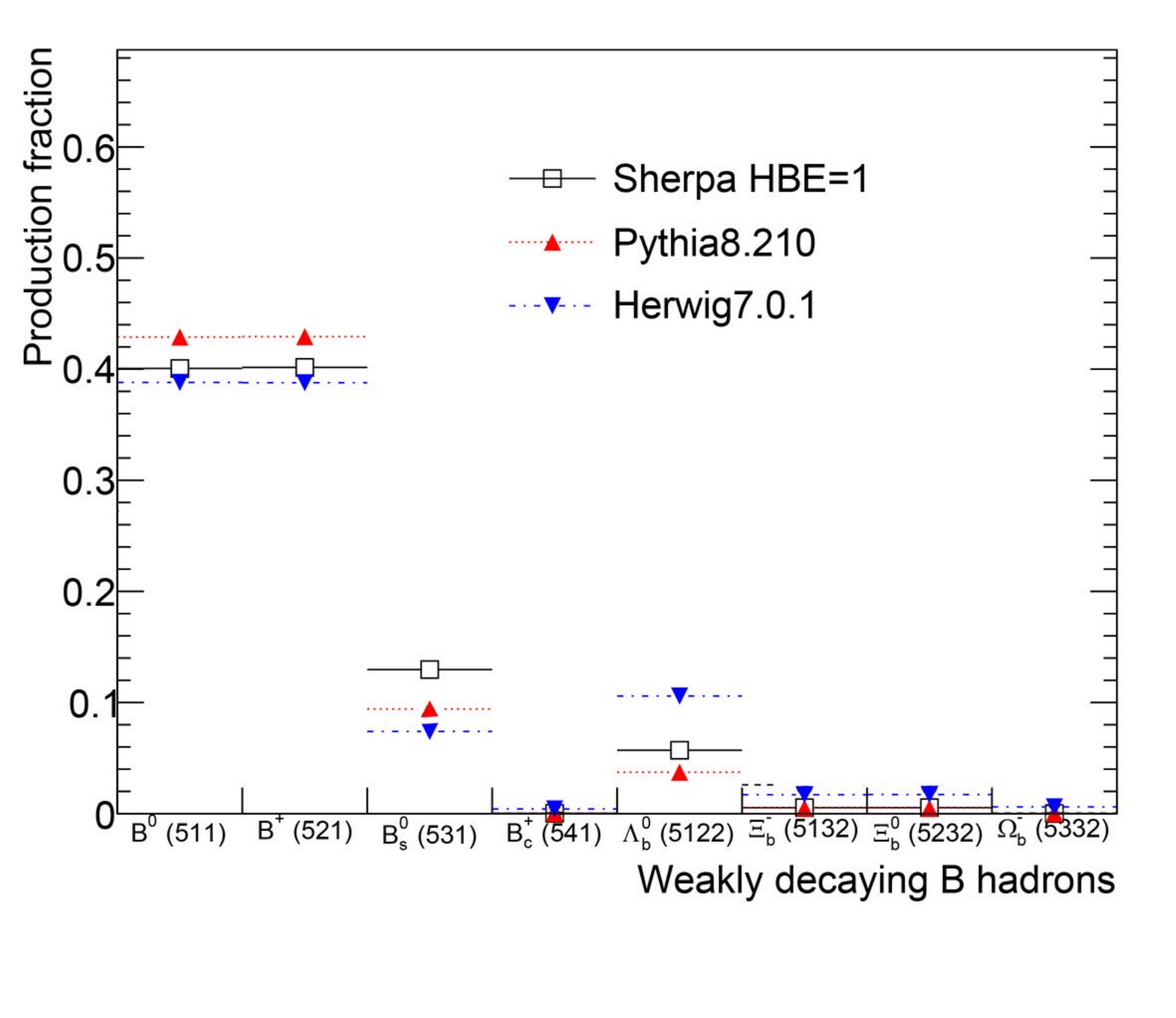}}
\caption{$b$-hadron production fractions at LHC extracted from several physics 
generators: Sherpa\,\cite{Gleisberg:2008ta}, Pythia\,\cite{Bahr:2008pv} and 
Herdwig\,\cite{Sjostrand:2007gs}. In the case of Sherpa HBE = 1 the Heavy Baryon Enhancement parameter (HBE) is set to one to improve the modelling of heavy quark hadronization. }
\label{fig:bhadron_prod}
\end{figure}

The fraction of $b$-baryon species produced in proton proton collisions at $\sqrt s $=1.96\;TeV is about $22\%$ \cite{PDG18}. Assuming that this proportion remains constant at $\sqrt s$ = 13 TeV, and that the total cross section $\sigma( {\rm pp}\to b{\bar b} X)$ is $\approx \;600~{\rm \mu b}$ \cite{Aaij:2016avz}, we get the values 
$\sigma( {\rm pp}\to \Lambda_b)^{13 {\rm TeV}} \sim 117\; {\rm \mu b}$ and $\sigma( {\rm pp}\to \Xi_b)^{13 {\rm TeV}} \sim  12\; {\rm \mu b} $. In this estimation 
the $\Omega_b^-$ contribution is taken to be about $10\%$ of the $\Xi_b^-$.
These values are inside the range of the predictions by several Monte Carlo generators, as shown in Fig. \ref{fig:bhadron_prod}.

\subsection{Fit procedure}
 
 An unbinned maximum likelihood fit is used to determine the photon and $b$-baryon polarization. The angular distribution for $\Lambda_{b}^{0}$ in Sec. \ref{subsec:Lb} allows to extract the photon polarization fixing the $\Lambda_{b}^{0}$ polarization, or extracting both parameters at the same time. Thus, Eq. \ref{eq:Lb_AngDist} is used for obtaining these two polarizations simultaneously. The dependence on the initial $b$-baryon polarization can be eliminated by integrating out the $\cos\theta_\Lambda$ in Eq. \ref{eq:Lb_AngDist} obtaining:

\begin{equation}\label{eq:Lb_AngDist_simpler}
 W(\theta_{\gamma},\theta_{p}) \propto  1-\alpha_{\gamma}\alpha_{\Lambda} \cos{\theta_p}.
\end{equation}

In the case of $\Xi_b^-$ angular distribution of Sec. \ref{subsec:Xib}, it is possible to integrate out the azimuthal angle $\eta$ in Eq. \ref{eq:Xi_AngDist}. This removes the dependence on the unknown $z_\Xi$ parameter:

\begin{flushleft}
\begin{align}\label{eq:Xi_AngDist_noAzi}
& W(\theta_{\Lambda},\theta_{p},\theta_{\Xi})\propto 1+ 
\alpha_{\Lambda}\alpha_{\Xi}\cos \theta_{p} +\alpha_{\gamma}\alpha_{\Xi}\cos 
\theta_{\Lambda} \nonumber \\  & +\alpha_{\Lambda}\alpha_{\gamma}\cos \theta_{p} \cos 
\theta_{\Lambda} - P_{\Xi_{b}}\alpha_{\Xi}\cos \theta_{\Xi} 
\cos \theta_{\Lambda} \nonumber \\ 
&  -P_{\Xi_{b}}\alpha_{\gamma}\cos \theta_{\Xi} -P_{\Xi_{b}}\alpha_{\Xi}\alpha_{\Lambda}\alpha_{\gamma}\cos \theta_{\Xi} \cos 
\theta_{p} \\  
&  -P_{\Xi_{b}}\alpha_{\Lambda}\cos \theta_{\Xi} \cos \theta_{\Lambda} \cos 
\theta_{p}. \nonumber
\end{align}
\end{flushleft}

This expression is used in the fit to extract the photon and $\Xi_b^-$ polarization simultaneously. 
Integrating out the $\cos \theta_\Xi$ angle in Eq.~\ref{eq:Xi_AngDist_noAzi}, the dependence with the $b$-baryon polarization can also be removed: 
 
 \begin{flushleft}
\begin{align}\label{eq:Xi_AngDist_simpler}
 W(\theta_{\Lambda},\theta_{p}) & \propto 1+ 
\alpha_{\Lambda}\alpha_{\Xi}\cos \theta_{p} +\alpha_{\gamma}\alpha_{\Xi}\cos \theta_{\Lambda}  \\  & +\alpha_{\Lambda}\alpha_{\gamma}\cos \theta_{p} \cos \theta_{\Lambda}. \nonumber
\end{align}
\end{flushleft}

In order to validate the fit procedure, to check consistency between pseudo-experiments and to detect possible biases in the distribution of the fitted parameters, pull distributions, defined as the difference between the fitted and generated values, divided by the uncertainty from the fit, are determined. It is found that there is no bias in the fit and the pull distributions are correct. 

\subsection{Measurement of $\alpha_\gamma$, $P_{\Lambda_{b}}$ and $P_{\Xi_{b}}$} \label{subsec:stats}
\begin{figure}
\centering
\resizebox{0.45\textwidth}{!}{\includegraphics{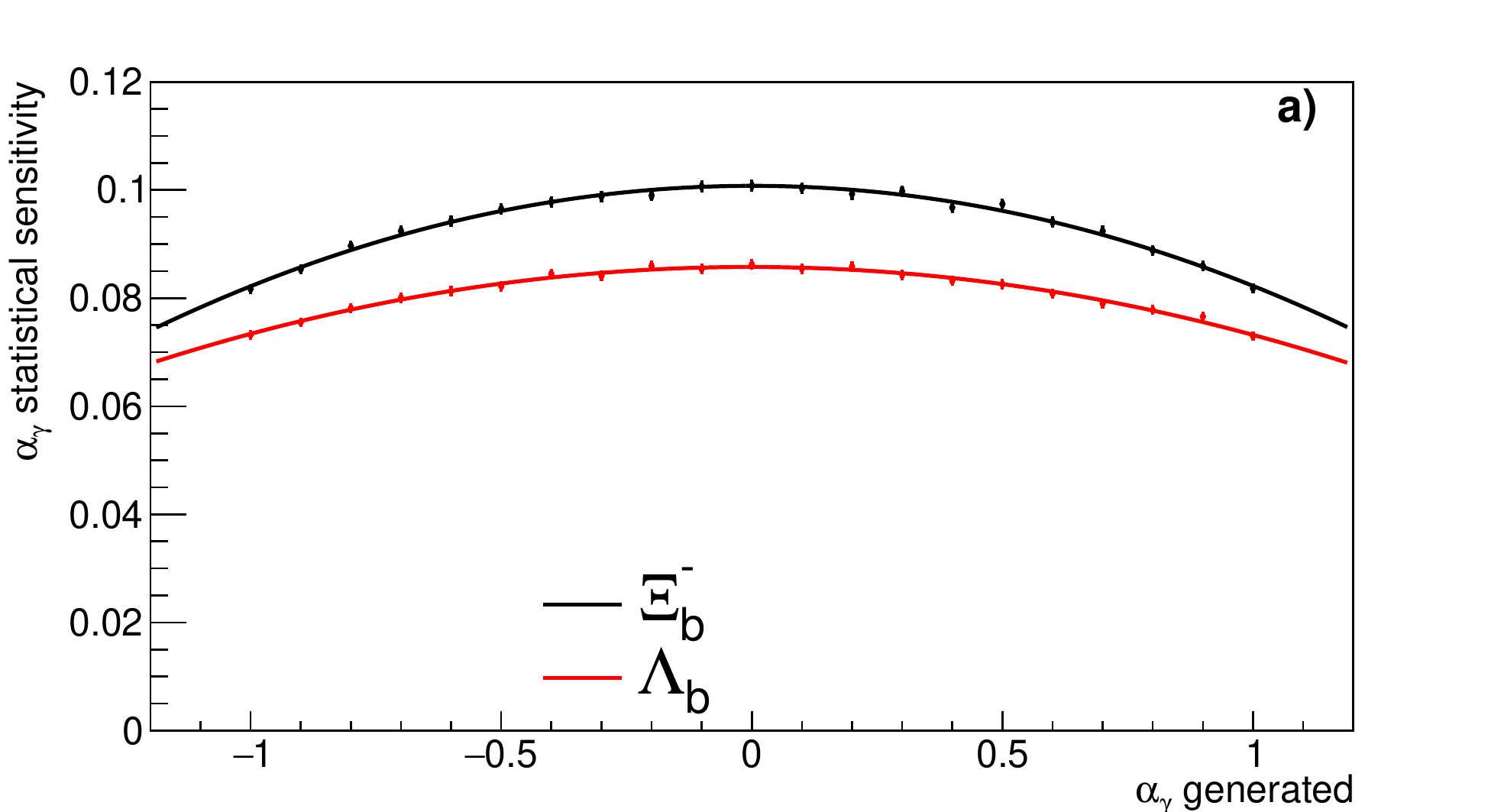}}
\resizebox{0.45\textwidth}{!}{\includegraphics{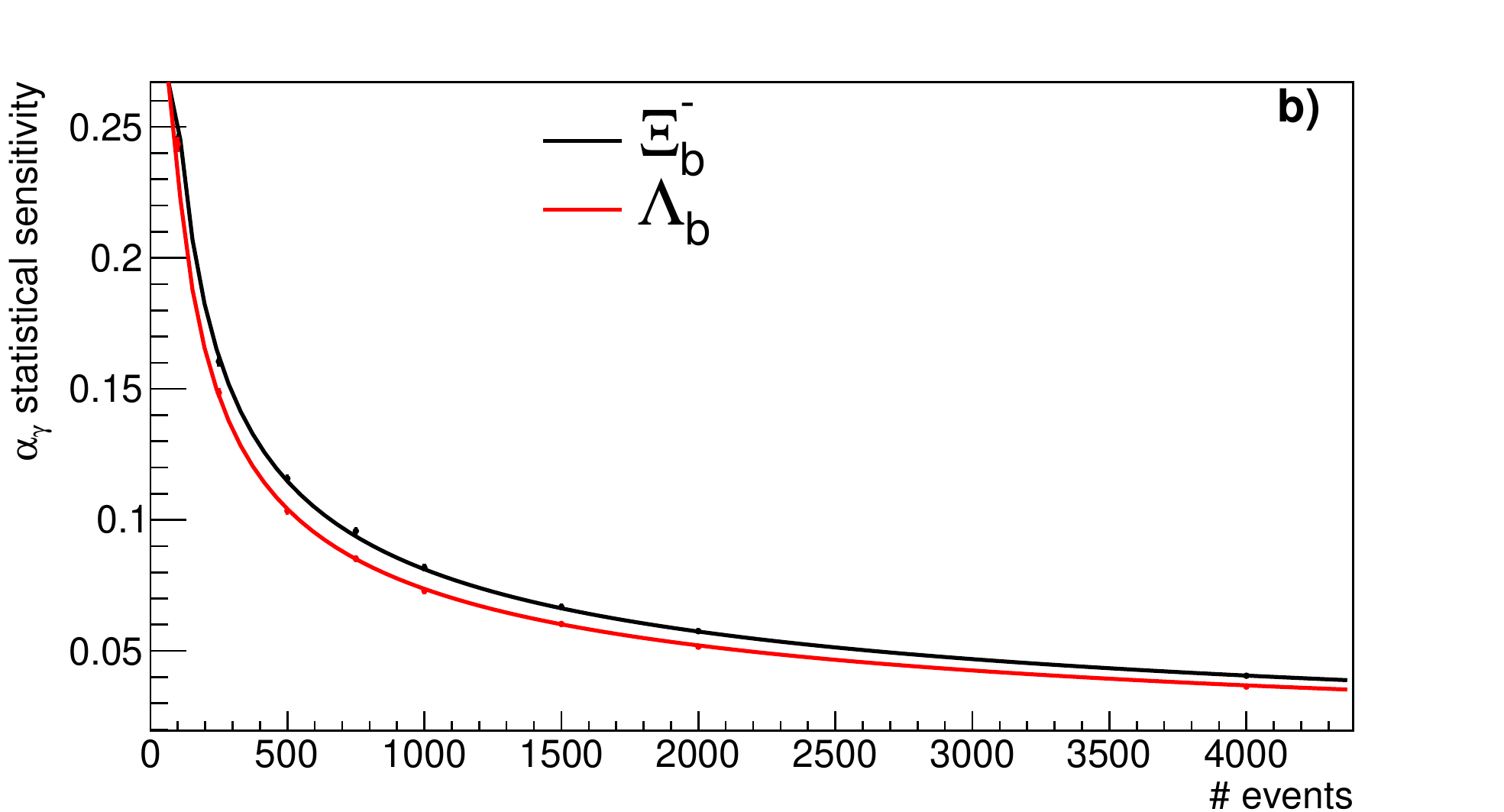}}
\caption{a) Sensitivity to the photon polarization parameter as function of its value for the $\Lambda_b^0$ and $\Xi_b^-$ decay channels. Dots with errors correspond to the fit results. A second order polynomial is superimposed for better visualization.
b) Statistical sensitivity of the photon polarization as function of the reconstructed number of events for the $\Lambda_b^0$ and $\Xi_b^-$ cases. In this case was fitted to a function $1/\sqrt{x}$.}
\label{fig:alfaSens_1000ev}
\end{figure}

To understand the correlation of the value of the $\alpha_\gamma$ parameter and its uncertainty, a large number of pseudo-experiments of 1000 events each are generated according to Eq.~\ref{eq:Lb_AngDist} and \ref{eq:Xi_AngDist} for the  $\Lambda_{b}^{0}$ and $\Xi_b^-$ decay channels respectively. The generated value of $\alpha_\gamma$ is varied in this study between $-$1.0 and +1.0 and the angular distribution is fitted for each case. The results of the fits are shown in Fig.~\ref{fig:alfaSens_1000ev} a). A small dependence with the value of the photon polarization is observed for both channels, with maximum sensitivity at larger slopes, +1 and $-$1 values of the $\alpha_\gamma$ parameter. The sensitivity for the $\Lambda_{b}^{0} \to \Lambda^0 \,\gamma$ decay channel is found to be slightly better (about $15\%$) as compared to the $\Xi_b^- \to \Xi^- \, \gamma$ channel. This is coming from the larger absolute value of the weak decay parameter, $\alpha_\Lambda$, compared to the $\alpha_\Xi$ parameter (see Eq. \ref{eq:Lb_AngDist_simpler} and \ref{eq:Xi_AngDist_simpler}). 
Fig. \ref{fig:alfaSens_1000ev} b) shows the expected sensitivity to the $\alpha_\gamma$ parameter as function of the reconstructed number of events for both channels. Pseudo-experiments with samples ranging between 100 and 4000 events are generated in this study with the Standard Model expectation of $\alpha_\gamma$= 1. 
The results show that the photon polarization can be potentially measured with less than $10\%$ uncertainty with the reconstruction of about 1000 radiative $b$-baryon events.  
\begin{table}
\centering
\caption{Expected signal yield for $\Lambda_b^{0} \to \Lambda^{0} \gamma$ and $\Xi_b^{-} \to \Xi^{-} \gamma$ decay channels.}\label{tab:stat}
\begin{tabular}{lll}
\hline \noalign{\smallskip}
& $\Lambda^{0}_b \rightarrow \Lambda^{0}\gamma$ & $\Xi^-_b \rightarrow \Xi^{-}\gamma$\\ 
   \noalign{\smallskip} \hline 
\noalign{\smallskip}
 $\sigma({\rm pp} \rightarrow H_b)^{13TeV} (\mu b)$ & $117$ & $12$ \\
 $BR(H_b \rightarrow H\gamma)$ & $10^{-5}$ & $10^{-5}$ \\
 $BR(\Xi^- \rightarrow \Lambda^{0}\pi^{-})$ & - & $1.0$ \\
 $BR(\Lambda^{0} \rightarrow p^{+}\pi^{-})$ & $0.64$ & $0.64$ \\
 $\epsilon_{H_b \rightarrow H_\gamma}$ & $1.0 \times 10^{-4}$ & $1.0 \times 10^{-4}$ \\
 \hline
 \noalign{\smallskip}
 $N~{\rm (Run\,II,~ 6\,fb^{-1})}$ & $900$ & $92$\\
 \hline
 \noalign{\smallskip}
 $N~{\rm (Run\,III,~ 25\,fb^{-1})}$ & $3740$ & $384$ \\
\noalign{\smallskip}\hline
\end{tabular}
\end{table}

In Table~\ref{tab:stat} the expected number of events for the different run periods at LHCb are estimated. Numbers are obtained according to the production fractions of the different $b$-baryon species (see Section~\ref{sec:fragmentation}) and assuming that the branching fraction for radiative $b$-baryon decays is about $10^{-5}$ \cite{Wang:2008sm}. A reconstruction efficiency of order $10^{-4}$ is considered. This estimation is obtained by scaling the reconstruction efficiency for radiative $B$ meson decays in \cite{Aaij:2016of} a factor 1/100. This assumption takes into account that the trigger and reconstruction algorithms at LHCb are less efficient for long living particles~\cite{PadLong}. Besides, one expects the need of tighter selection criteria to suppress the large amount of combinatorial background expected from prompt $\Lambda^0$'s and random photons.     

\begin{figure}
\centering
\resizebox{0.45\textwidth}{!}{\includegraphics{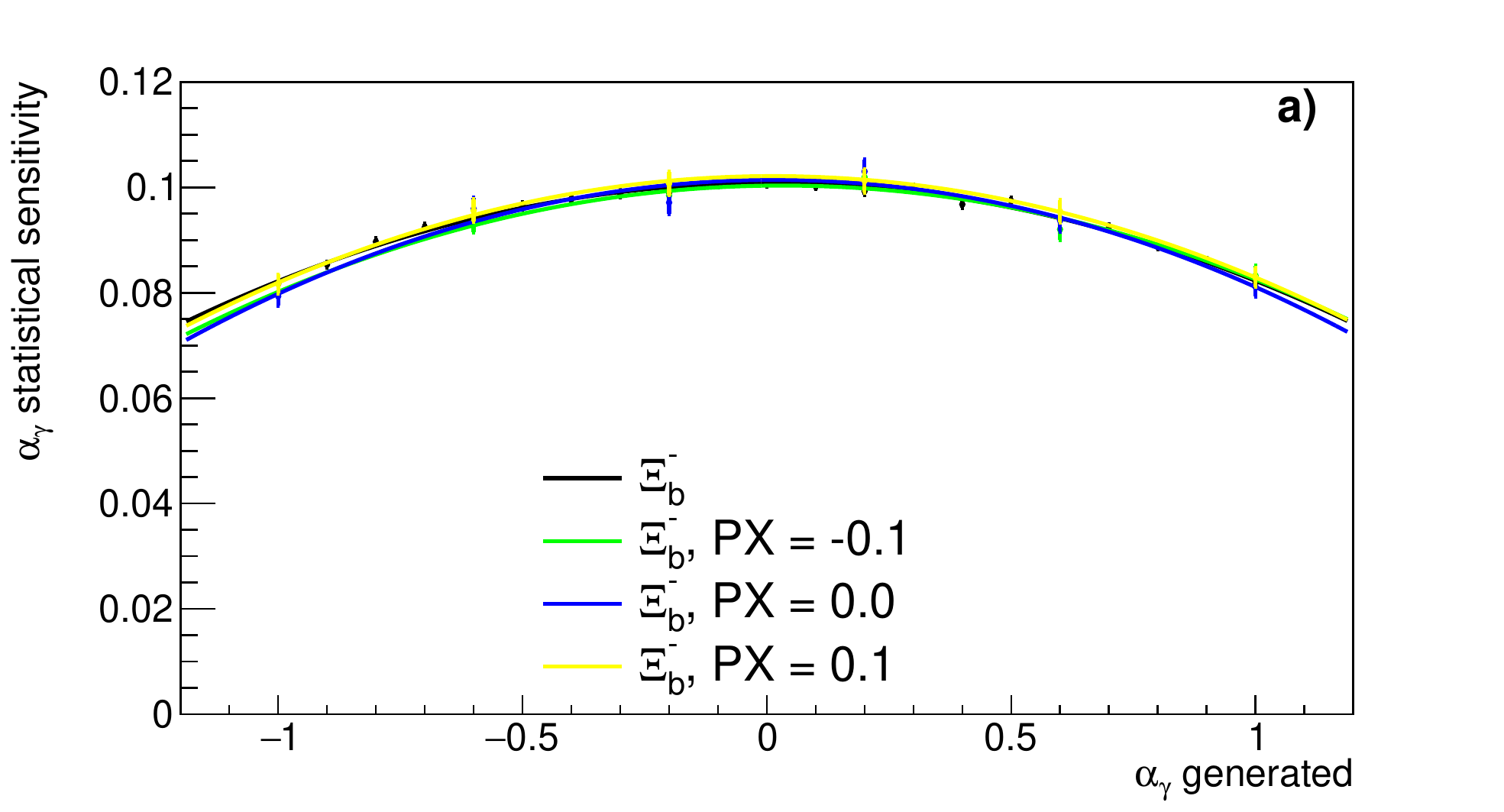}}
\resizebox{0.45\textwidth}{!}{\includegraphics{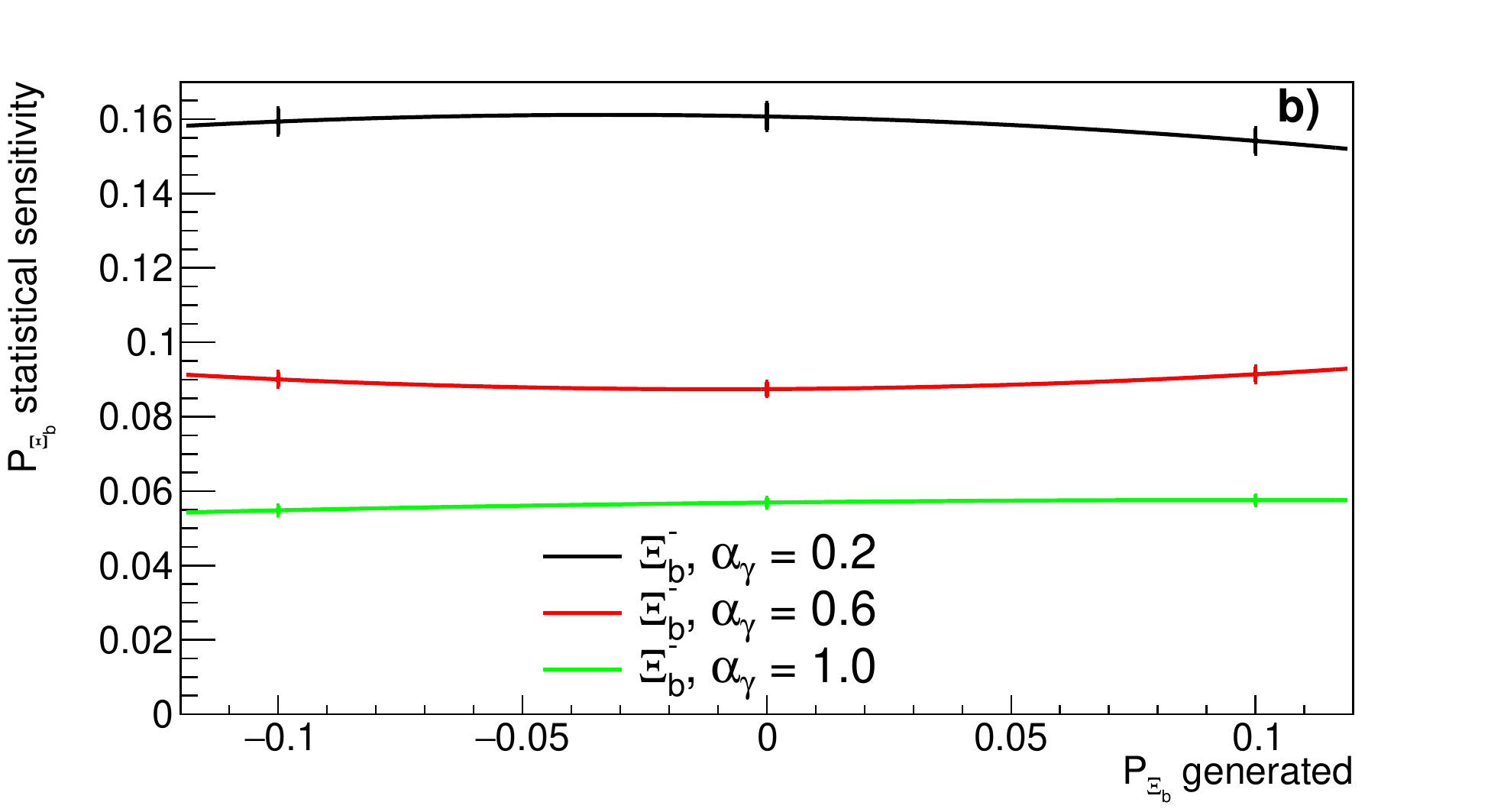}}
\resizebox{0.45\textwidth}{!}{\includegraphics{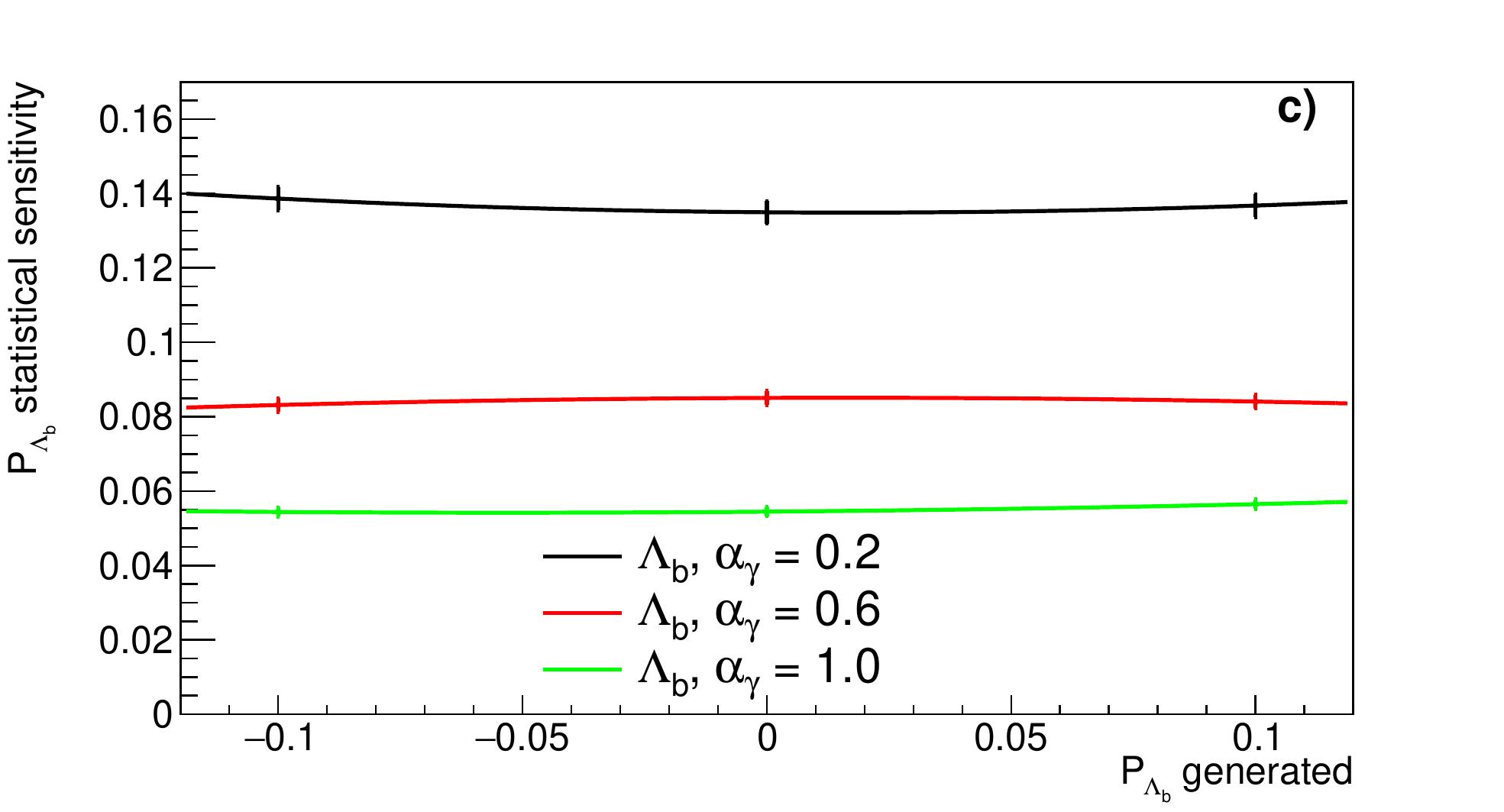}}
\caption{a) Sensitivity to the photon polarization as function of its value for different assumptions of the initial $\Xi_b^-$ polarization. 
The black line corresponds to the fit of Eq. \ref{eq:Xi_AngDist_simpler}, which does not include the polar angle of the $\Xi^-$ momentum relative the normal vector to the production plane, while the other line correspond to the the fit of Eq. \ref{eq:Xi_AngDist_noAzi}. Similar results are obtained for the $\Lambda_b^0$ polarization. 
b) Sensitivity to the ${\Xi_b^-}$  polarization as function of its value for $\Xi_b^-$ for different values of $\alpha_\gamma$ using Eq. \ref{eq:Xi_AngDist_noAzi}.
c) Sensitivity to the ${\Lambda_b^0}$  polarization as function of its value for $\Lambda_b^0$ for different values of $\alpha_\gamma$ using Eq. \ref{eq:Xi_AngDist_noAzi}. 
In these figures dots with errors correspond to the fit results. A second order polynomial is superimposed for better visualization.}
\label{fig:PolSens}
\end{figure}

Since the angular distribution of radiative $b$-baryon decays is also sensitive to the initial $b$-baryon polarization (see Eq. \ref{eq:Lb_AngDist} and \ref{eq:Xi_AngDist_noAzi}), one can extract $P_{\Lambda_{b}}$ and $P_{\Xi_{b}}$ values together with the photon polarization. Pseudo-experiments with 1000 events each are performed in this study using several values of the $\Lambda_{b}^0$ and $\Xi_{b}^-$ polarization, fitting simultaneously $\alpha_\gamma$ and $P_{\Lambda_{b}}$, for the 
$\Lambda_{b}^{0} \to \Lambda^0 \, \gamma$ decay channel, and $\alpha_\gamma$ and $P_{\Xi_b}$ for the $\Xi_{b}^{-} \to \Xi^{-}\, \gamma$ decay channel. Since the $P_{\Xi_{b}}$ in hadron collisions is expected to be similar to the polarization of the $\Lambda_b$, which has been measured in \cite{Sirunyan:2018bfd}, a maximum polarization of $10\%$ for the initial $b$-baryon polarization is considered in the generation for both channels. The results of the achieved sensitivity for the $\alpha_\gamma$ and $P_{\Lambda_{b}}$ or $P_{\Xi_{b}}$  parameters as a function of its value is shown in Fig.  \ref{fig:PolSens}. 

As it can be seen in Fig. \ref{fig:PolSens} b), the sensitivity to the $\alpha_\gamma$ parameter is not affected by the measurement of the $\Xi_{b}^{-}$ polarization in the simultaneous fit, given enough statistics. This is also true for the same study with the $\Lambda_{b}^{0}$ polarization.
In addition, the value of the $\Lambda_{b}^{0}$ and $\Xi_{b}^{-}$ polarization parameters can be extracted with good accuracy independently of its value, with improved precision 
as the photon polarization approaches the SM value.

\subsection{Event reconstruction and background effects}\label{subsec:reco}

\begin{figure}
\centering
\resizebox{0.45\textwidth}{!}{\includegraphics{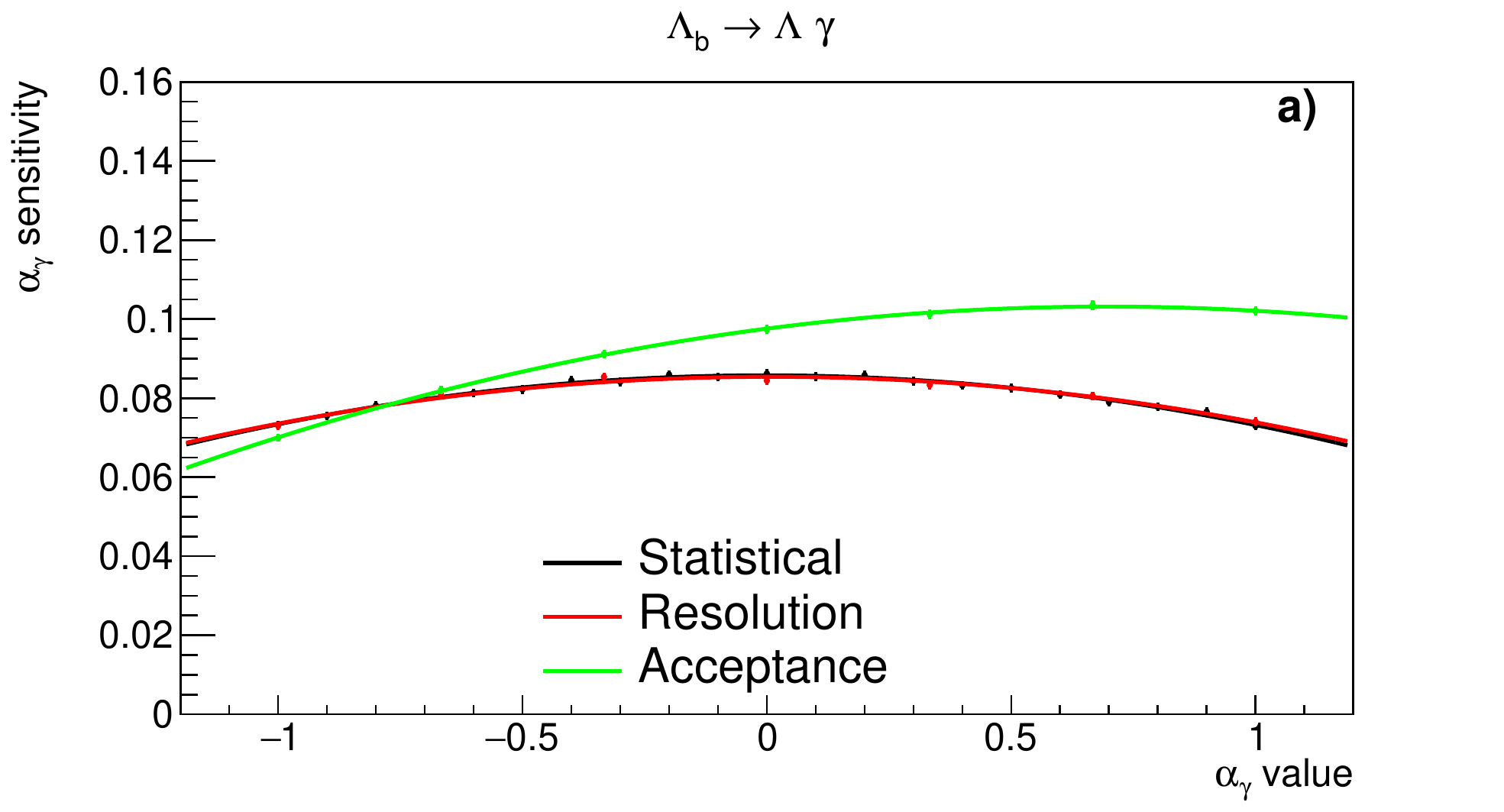}}
\resizebox{0.45\textwidth}{!}{\includegraphics{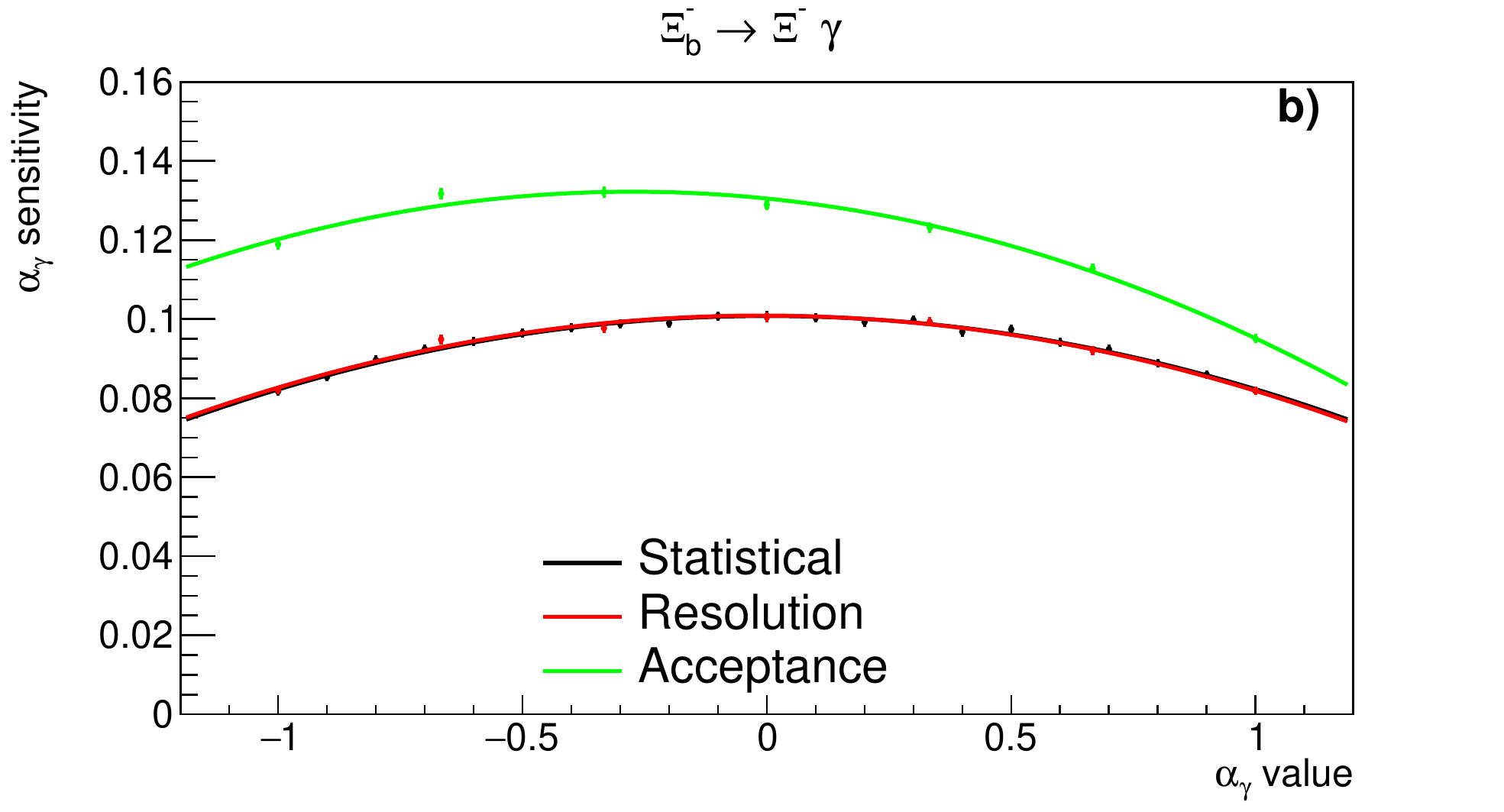}}
\caption{a) Sensitivity to the photon polarization as function of its value for $\Lambda_{b}^{0}$ (a) and $\Xi_b^-$ (b) cases including angular resolution and acceptance effects. 
Dots with errors corresponds to the fit results. A second order polynomial is superimposed for better visualization.}
\label{fig:Lb1000}
\end{figure}

To include the effect of the signal event reconstruction and background sources the angular distribution in Eq.~\ref{eq:Lb_AngDist_simpler} and Eq.~\ref{eq:Xi_AngDist_simpler} is modified. The signal probability density distribution is affected by an angular acceptance, ${\cal A}(\theta_\Lambda, \theta_{p})$, and a resolution function, 
${\cal R}(\theta_\Lambda, \theta_{p}; \theta_\Lambda', \theta'_{p} )$. Additionally, background sources with different angular shapes 
$f_{B}$ and signal to background rate ($S/B$) are also considered in the fitting procedure. The angular distributions of the $\Lambda_b^0$ and $\Xi_b^-$ decay channels are modified:       

\begin{multline*}
\nonumber
\Gamma (\theta'_\Lambda, \theta'_{p}; \alpha_\gamma) =  \\ 
\nonumber
\Big(f_{S}(\theta_\Lambda, \theta_{p}; \alpha_\gamma) \times
{\cal A}(\theta_\Lambda, \theta_{p}) \Big) \ast {\cal R}(\theta_\Lambda, \theta_{p}; 
\theta'_\Lambda, \theta'_{p}) + \\
\frac{S}{B}\big(f_{B}(\theta_\Lambda, \theta_{p})\big) 
\end{multline*}
The LHCb detector has been described elsewhere \cite{Aaij:2014jba,Alves:2008zz}.
 The resolution of the proton and $\Lambda^0$ polar angles are expected to be worse if the $\Lambda^0$ particle decays after the first LHCb tracker, since the vertex and momentum resolutions degrade. If the information of the LHCb VELO detector is not included in the reconstruction of the proton and pion candidates, named {\it downstream} track category, the angular resolution can be as large as 90 mrad \cite{Aaij:2018gwm}. Since $\Lambda^0$ particles have a long lifetime, most of the events are expected in this category. No bias in the mean of the angular resolution is expected.   
 For tracks which are reconstructed including the information of the VELO detector,  named {\it long} track category, the angular resolution is expected to be better. Monte Carlo simulations are performed using angular resolutions from 30 to 90 mrad.  Since the angular distributions are quite smooth the resolution is expected to have a negligible effect. Figure \ref{fig:Lb1000} shows the effect of the resolution and angular acceptance for the $\Lambda_{b}^{0}$ and $\Xi_{b}^{-}$ cases. 

The angular acceptance is expected to be due to trigger and reconstruction selection criteria. The most biasing effects are expected to come from pion momentum requirements, affecting the proton and $\Lambda^0$ angles in an asymmetric way. Following the results obtained in \cite{Aaij:2013oxa} for the proton angle, a third order polynomial function is implemented for the acceptance. Pseudo-experiments are generated using this function. Results of the fits, including the acceptance, are shown in Fig.~\ref{fig:Lb1000}. 
An important effect on the sensitivity to the photon polarization is observed, which will have to be controlled using data control samples. For the case of the $\Xi_b^-$ small correlation between the proton and $\Lambda^0$ angles is expected.   

\begin{figure}
\centering
\resizebox{0.45\textwidth}{!}{\includegraphics{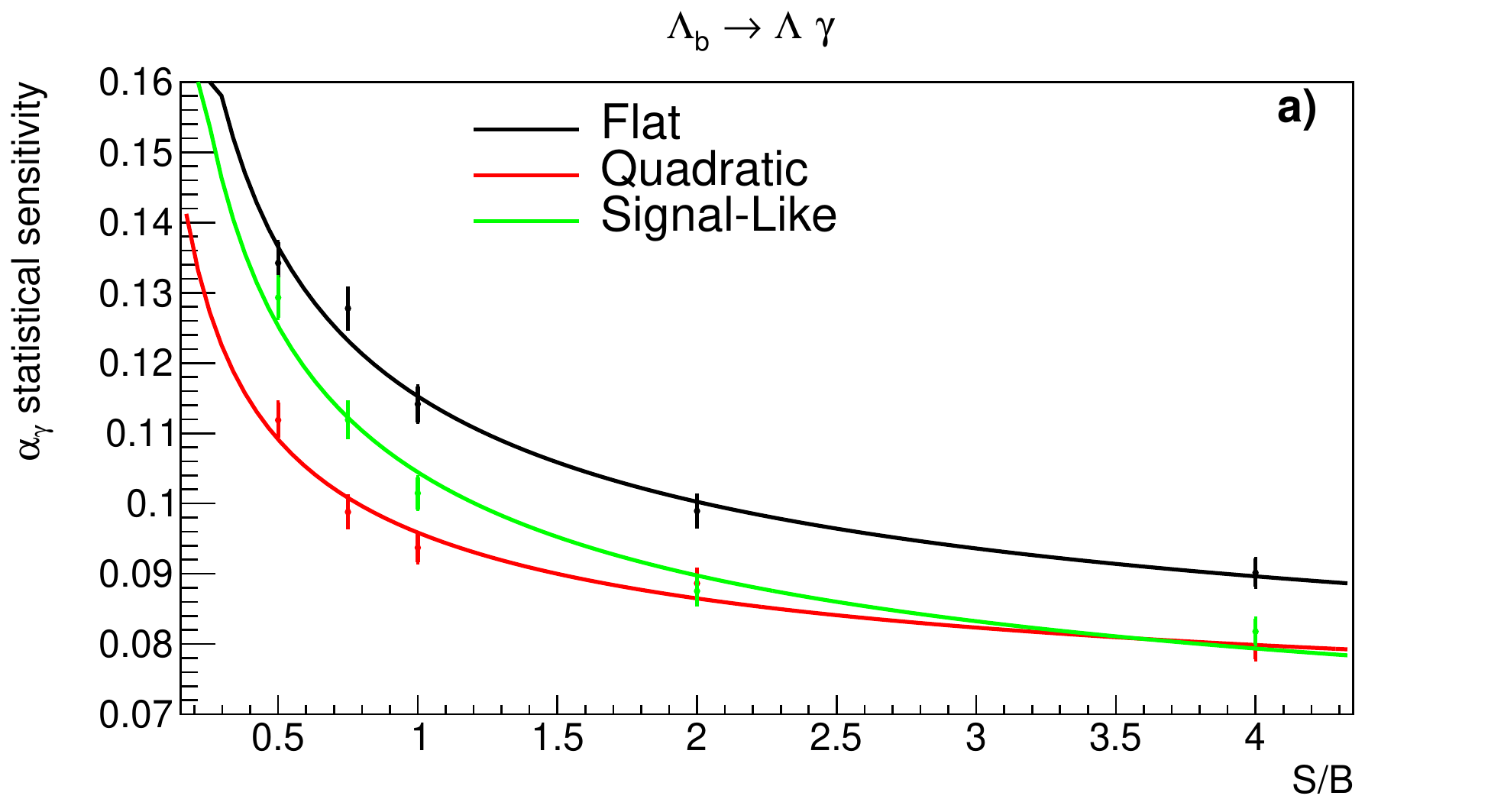}}
\resizebox{0.45\textwidth}{!}{\includegraphics{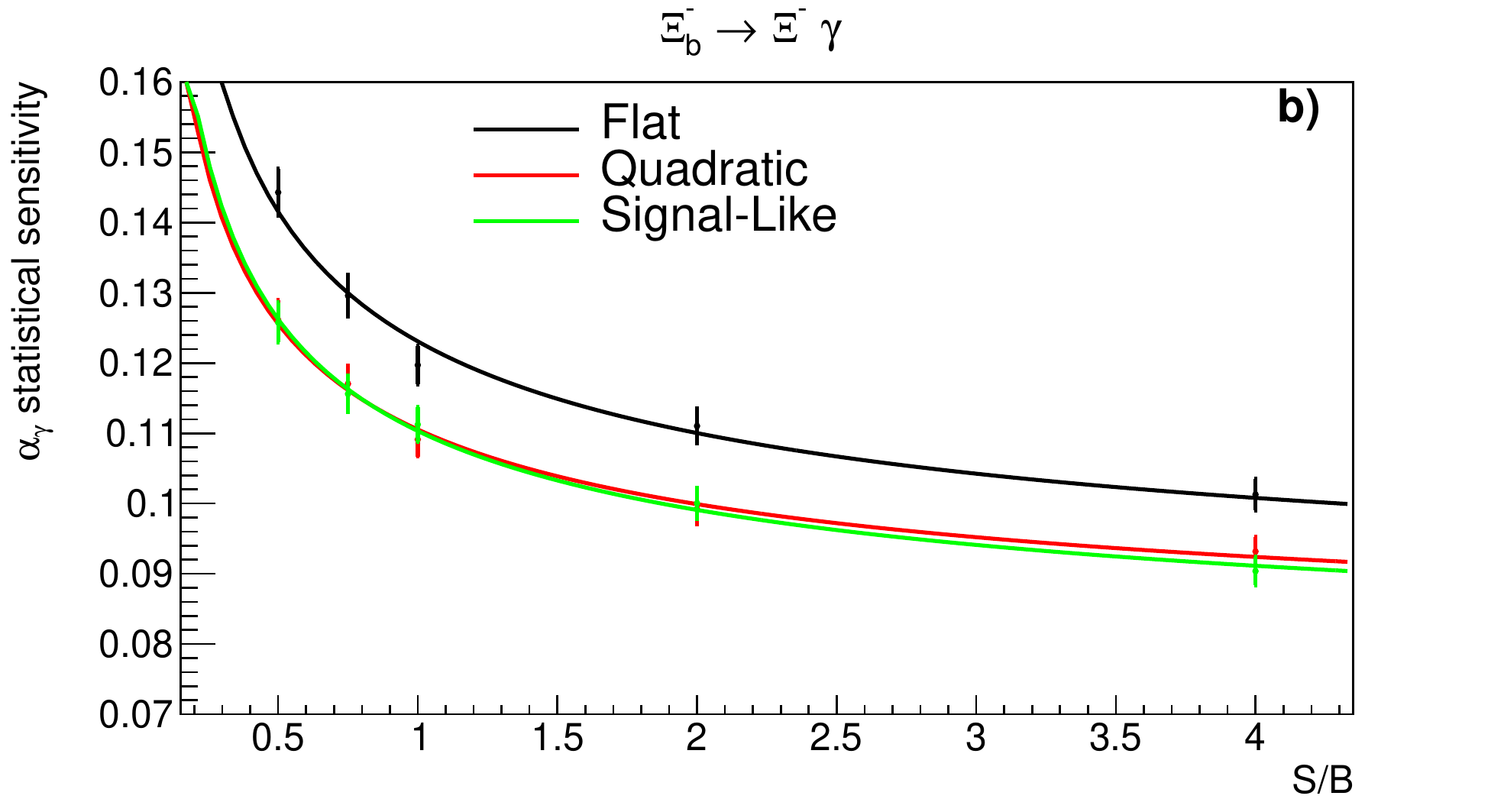}}

\caption{Sensitivity to the photon polarization as function of the signal over background ratio for $\Lambda_b^{0}$ (a) and  $\Xi_b^-$ (b) including background effects. 
Dots with errors corresponds to the fit results. A function $A + B/\sqrt{x}$ is superimposed for better visualization.}
\label{fig:XiB_BKGSens}
\end{figure}

One of the most important challenges when reconstructing radiative $b$-baryon decays is expected to be the background suppression. Since strange $b$-baryons have large lifetimes, and photons are involved in the decays, making a vertex fit from the $b$-baryon decay products is not possible.
A large amount of combinatorial background is then expected from random photons and prompt $\Lambda^0$ particles. In this study a background source is considered and a large number of pseudo-experiments is generated and fitted considering different signal (S) to background (B) rates, as well as different shapes for the angular distribution of the background. The signal samples are of 1000 events each, and the amount of events in the background sample are varied. A flat, signal-like and a second order polynomial are tested for the angular distribution of the background. The samples are generated with SM values of the photon polarization parameter, $\alpha_\gamma = 1$. Fig. \ref{fig:XiB_BKGSens} shows the results of the fit of the pseudo-experiments.

As it can be observed, the shape of the background is not so relevant as the background rejection capability. An important S/B rate is necessary in order to measure the photon polarization with an accuracy below $10\%$. To achieve this, multivariate analysis techniques will be crucial, in particular for rejecting the large amount of the expected combinatorial background.

\subsection{Discussion of results} 
Assuming that the LHCb detector can reconstruct about 900 $\Lambda_b^0 \to \Lambda^0 \, \gamma$ events with the data of the Run II, and that systematic uncertainties can be controlled to the same order of the statistical uncertainty, a sensitivity of about 0.15 could be achieved for the photon polarization parameter $\alpha_\gamma$.
The $\Lambda_b^0$ polarization can also be measured with a precision better than $10\%$, the sensitivity depending on the value of the photon polarization. 

The measurement for the $\Xi_b^- \to \Xi^- \, \gamma$ decay channel is more challenging and will need more data to reach the same sensitivity but it is a promising channel for Run III and beyond. It has the advantage of an additional track coming from the $\Xi^-$ ($\to \Lambda^0 \, \pi^-$) and the possibility to reconstruct the charged $\Xi^-$ baryon. If systematic uncertainties are controlled at the order of the statistical uncertainty, a sensitivity of around 0.2 could be achieved.  
This channel also offers the possibility to measure the $\Xi_b^-$ polarization, which is unknown at present. The experimental angular resolution is found to have a negligible effect on the measurement of the photon polarization. In contrast, the effect of the angular acceptance due to the event reconstruction is found to be important needs to be controlled from data. 

The crucial task in these analyses is expected to be the background suppression. A signal-to-background rate less than one is needed to obtain a good precision in the photon polarization.   

Estimations in Tab.~\ref{tab:stat} for the LHCb Run III do not assume any improvement of the LHCb detector, which is far from reality. In the next LHC run, from 2020-2023, in addition to upgrade its tracking and vertexing systems, the LHCb experiment will exploit a novel trigger concept where all subdetectors are read out in real time and the first level trigger is fully software implemented. Track momenta resolution will improve by 10-20$\%~$\cite{upgrades}. For long living particles such as strange baryons, the efficiency is expected to increase at least by 10$\%$ using dedicated reconstruction algorithms
\cite{PadLong}, and could still be much larger with the improvement of the software trigger.  

\subsection{Normalization channel}
\label{sec:norma}
A data-driven measurement of radiative b-baryon decays require the use of control and normalization decay modes.   
The $\Lambda_{b}^{0} \to \Lambda^0 \, J/\psi$ $ (\to \mu^+\,\mu^-)$ and $\Xi_{b}^{-} \to \Xi^- \,J/\psi (\to \mu^+\,\mu^-)$ are chosen as the preferred modes of normalization for the $\Lambda_{b}^0 \to \Lambda^0 \, \gamma$ and $\Xi_{b}^- \to \Xi^- \, \gamma$ channels, respectively. The advantage of this choice is that most of the systematic uncertainties coming from the track reconstruction of $b$-baryons cancel when the ratio of rates are considered. Also the fragmentation fractions, which have at present large uncertainties (see Sec. \ref{sec:fragmentation}), cancel. The relative number of events of the radiative and di-muonic channels is then:  

\begin{eqnarray}
\frac{N_{\Lambda_{b}^0\to \Lambda^0 \, \gamma}}{N_{\Lambda_{b}^{0}\to \Lambda^0 \, J/\psi}} = \frac{\mathcal{B} (\Lambda_{b}^0 \to \Lambda^0 \,\gamma) \, \epsilon_{\Lambda^0 \, \gamma} }
{\mathcal{B} (\Lambda_{b}^0 \to \Lambda^0 \, J/\psi) \mathcal{B}(J/\psi\to  \mu^{+}\, \mu^{-}) \,\epsilon_{\Lambda^0\, J/\psi}}, \nonumber
\end{eqnarray}
assuming the same luminosity for both channels, and taking into account that $f_{\Lambda_{b}}$, and $\mathcal{B}(\Lambda^0\to p\,\pi^-)$ cancel in this ratio.
$\epsilon_{\Lambda^0 \, \gamma}$ and $\epsilon_{\Lambda^0 \, J/\psi}$ are the remaining efficiencies which are different for both channels, coming mainly from photon and muon reconstruction. In a similar way: 
\begin{eqnarray}
\frac{N_{\Xi_{b}^{-}\to \Xi^{-} \gamma}}
{N_{\Xi_{b}^{-} \to \Xi^- J/\psi}} = 
\frac{\mathcal{B}(\Xi_{b}^-\to \Xi^- \gamma) \, \epsilon_{\Xi^- \gamma}}
{\mathcal{B}(\Xi_{b}^-\to \Xi^- J/\psi)\mathcal{B}(J/\psi\to \mu^{+} \mu^{-}) \, \epsilon_{\Xi^- J/\psi}}, \nonumber
\end{eqnarray}
where again same luminosity for both channels is assumed, and $f_{\Xi_{b}}$, $\mathcal{B}(\Xi^-\to \Lambda^0 \, \pi^-)$ and $\mathcal{B}(\Lambda^0 \to p \, \pi^-)$ cancel. 
$\epsilon_{\Xi^- \,\gamma}$ and $\epsilon_{\Xi^- \, J/\psi}$ are the remaining efficiencies which are different for both channels. 
Note that no attempt is made to measure the angular distribution of di-muonic channels. 

An alternative decay channel for normalization is the $B^0\to K^{*0} \, \gamma$, which has larger efficiency but is also affected by larger amount of background. In this case the systematic uncertainty due to the photon reconstruction cancels.  

\section{New physics constraints}
\label{sec:flavio}
The process $b \to s \gamma$, at the leading order in $\alpha_{s}$, gets contribution from dimension six 
electromagnetic dipole operators:
$$O_{7}^{(\prime)}\,=(em_{b})/(16\pi^2)\bar{s}\sigma^{\alpha \beta}P_{R(L)}b\,
F_{\alpha \beta}$$ corresponding to the emission of a left (L) or right (R) handed photon
respectively. The small $s$-quark mass has been ignored in $O_{7}^{(\prime)}$ and $P_{R,L}$ are defined as $P_{R,L}=(1\pm\gamma_{5})/2$. The Hamiltonian corresponding to the decay is given by 
\begin{eqnarray}
 \mathcal{H}_{eff}= 
 \frac{-4 G_{F}\,V_{tb}V_{ts}^{*}}{\sqrt{2}}(C_{7}O_{7}+C_{7}^{'}O_{7}^{'})~,
\end{eqnarray}
where $C_{7}^{(\prime)}$ are the Wilson coefficients \cite{Mannel:1997xy} describing the short 
distance contributions, $V_{tb}$ and $V_{ts}$ are the relevant Cabibbo-Kobayashi-Maskawa (CKM) matrix elements for the transition, and $G_{F}$ is the Fermi constant. 
The polarization asymmetry $\alpha_\gamma$ in Table~\ref{tab:dec param} can be thus defined as 
\begin{eqnarray}
  \alpha_{\gamma}=\frac{\Gamma(\lambda_{\gamma}=L)-\Gamma(\lambda_{\gamma}=R)}
  {\Gamma(\lambda_{\gamma}=L)+\Gamma(\lambda_{\gamma}=R)},
\end{eqnarray}

\begin{figure}
\resizebox{0.45\textwidth}{!}{\includegraphics{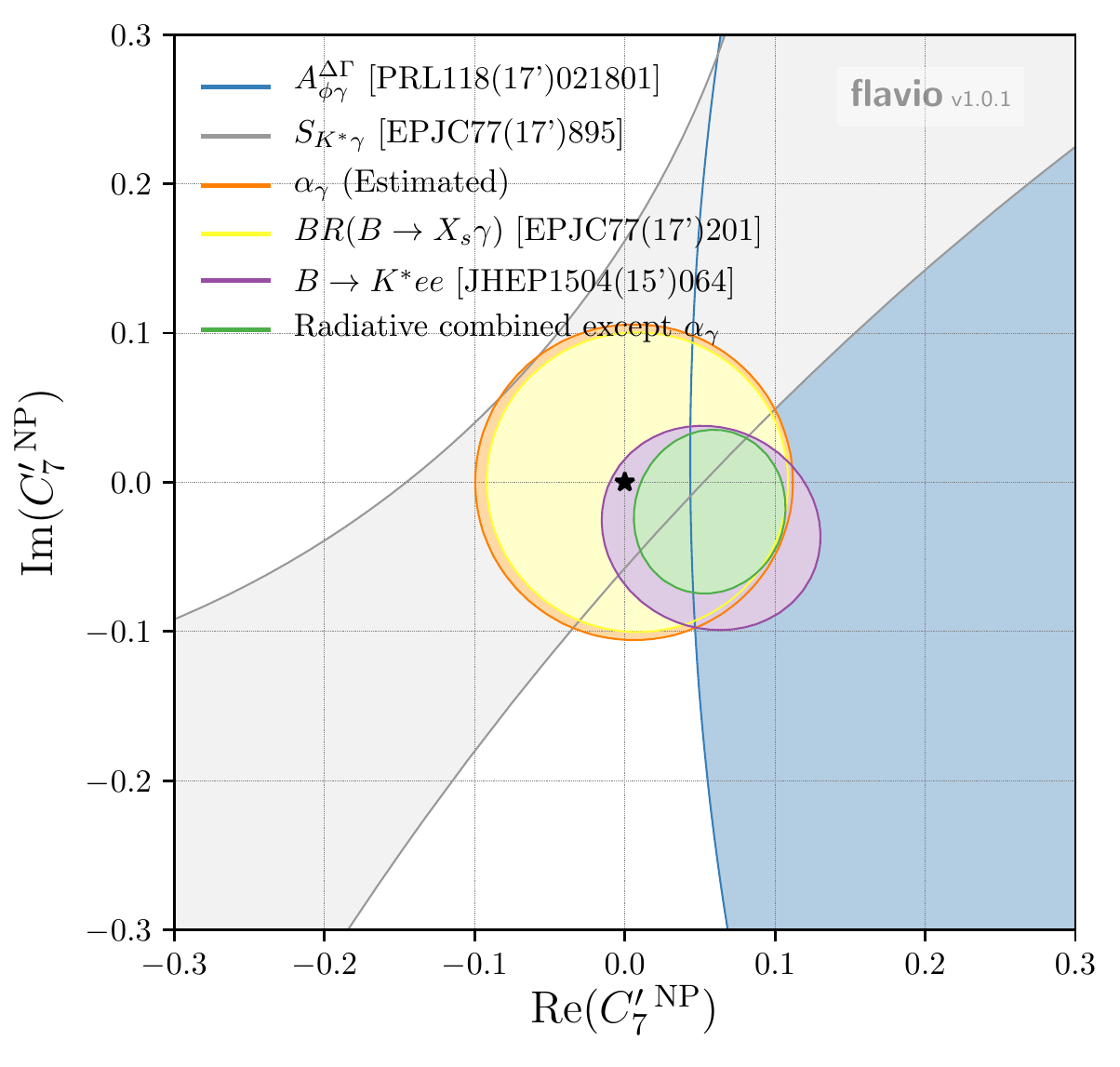}}
\caption{New Physics constraints on $\text{Re}(C_7')-\text{Im}(C_7')$ using flavio \cite{Paul:2016urs,Straub:2018kue}.  $A_{\phi\gamma}^{\Delta\Gamma}$ 
in blue, $S_{K^* \gamma}$ in violet, $\alpha_\gamma$ (assuming a value of $\alpha_\gamma 
= 1.0 \pm 0.15$) in orange, the inclusive radiative branching ratio in yellow, the angular study of $B\to K^* \,e^+\,e^-$ in purple and the combination of all this measurement except for the $\alpha_\gamma$ one in green. Besides, the SM value is marked with a star.}

\label{fig:flavio}
\end{figure}

Within SM the 
contribution to right-handed photons is suppressed by the ratio
$r=\frac{C_{7}^{'}}{C_{7}}\simeq\frac{m_{s}}{m_{b}}$, resulting in
\begin{equation}
\alpha_\gamma 
=\frac{1-\vert r  \vert^{2}}{1+\vert r \vert^{2}}\simeq 1.
\end{equation}
However, various new physics (NP) scenarios as well as long distance effects
within the SM can result in new contributions to the $C_{7}^{'}$ causing a
departure from the SM value \cite{Paul:2016urs}. Long distance contribution has been estimated to be small and of the order of 5\% \cite{Mannel:1997xy}. Thus, any significant deviation of the 
measured value of $\alpha_{\gamma}$ from unity could indicate the 
presence of NP contributions to the Wilson coefficient  $C_7^{\prime}$. In 
general $C_7^{\prime}$ can have both real and imaginary contributions. 
There are, however, already constraints on $C_7^{\prime}$ using other measurements like $A_\text{CP}$, the direct CP asymmetry, $S$, the mixing-induced CP asymmetry and $A_{\Delta\Gamma}$, the mass-eigenstate rate asymmetry from the decays of neutral $B_d$ and $B_s$ mesons to CP eigenstates,
for example in $B_s\to\phi\, \gamma$ or $B^0\to K^*(\to K_S\, \pi^0)\, \gamma$ decays.
In Fig.~\ref{fig:flavio}, the constraints in the $C_7^{\prime}$ plane from measurements of $A^{\Delta\Gamma}(B_s\to \phi \, \gamma)$, the inclusive branching ratio $Br(B\to
X_s\, \gamma)$ , $S(B\to  K^*\, \gamma )$ and angular asymmetries of $B\to K^*\,e^+\,e^-$ are shown.
The constraint for an estimate of $\alpha_{\gamma}=1.0 \pm 0.15$ is also plotted. With the expected sensitivity of $\alpha_{\gamma}$ it is clear that this measurement provides competitive constraints on new physics in comparison to the existing measurements using $B$ mesons. 

\section{Conclusions}
\label{sec:conc}
\vspace{0.5cm}

The flavour-changing GIM suppressed processes involving $b\to s \gamma$
transitions are known to be excellent probes of physics beyond the Standard
Model. In that context, we analyze the decays $\Lambda_{b}^{0}\to \Lambda^{0} \,\gamma$ and
$\Xi_b^{-} \to \Xi^{-} \gamma$ undergoing the same transition. By exploiting the spin correlations in radiative
decays of $b$-baryons we demonstrate that they offer a convenient method for
direct measurement of the  photon polarization asymmetry $\alpha_{\gamma}$ along with a measurement of the initial $b$-baryon polarization.
We  presented expressions for the complete
angular distribution of the decay chains $\Lambda_{b}^{0}\to \Lambda^0 \,\gamma$
and $\Xi_{b}^-\to \Xi^-(\to \Lambda^0\,\pi^-)\,\gamma$, with the $\Lambda^0$
decaying into $p\, \pi^-$, expressed as a function of $\alpha_{\gamma}$, the photon
polarization, $P_b$, the initial $b$-baryon polarization and known decay
asymmetry parameters of intermediate baryons. We then explored the potential of
joint measurement of the photon and $b$-baryon  polarization at LHCb
through the angular distribution study of $b$-baryon decays using Monte Carlo simulations. We find that with the expected yield from
the LHC Run II a sensitivity of about 0.15 is achievable for the photon
polarization along with a measurement of $\Lambda_{b}^{0}$ polarization with a
precision better than 10\%. Despite of the challenges in $\Xi_{b}^{-}$ due to
scarcity of data, even Run II allows a sensitivity of the order of 0.2 to be
achieved for $\alpha_{\gamma}$ along with a first time measurement of
$P_{\Xi_{b}}$. With higher yields, achievable in subsequent runs of the Large Hadron Collider, the photon polarization measurement using $b$-baryons will play a pivotal role in
constraining different new physics scenarios. 

\section*{Acknowledgements}
We thank L. Oliver for the interesting and motivating discussions at the beginning of this work. Part of this work is supported by the Spanish Ministerio of Econom\'ia, Industria and Competitividad.



\end{document}